\documentclass[epj]{svjour}
\usepackage{amssymb,amsmath}
\usepackage{graphics}
\usepackage{array}
\usepackage{wrapfig}
\begin{document}
\title{Wounded Quarks and Multiplicity at Relativistic Ion Colliders}
\author{Ashwini Kumar\inst{1}\thanks{\emph{e-mail : ashwini.physics@gmail.com}} \and B. K. Singh \inst{1} \and P. K. Srivastava\inst{1}\thanks{\emph{e-mail : prasu111@gmail.com}} \and C. P. Singh\inst{1}
}                     
%
%
\institute{Department of Physics, Banaras Hindu University, Varanasi 221005, INDIA}
\date{Received: date / Revised version: date}
%
\abstract{
In this paper, we propose a parameterization which is based on a phenomenological model involving the wounded quarks interactions for explaining the average charged particle multiplicity $\langle n_{ch}\rangle$, the central pseudo-rapidity density $\langle (dn/d\eta)_{\eta=0}\rangle$ and complete rapidity dependence of $dn/d{\eta}$ in relativistic heavy-ion collider experiments. The model also interrelates nucleus-nucleus (A-A) collisions with p-A and p-p interactions. Our parameterization rests on simple assumptions regarding mean number of participating quarks and their average number of collisions. The results for $\langle n_{ch}\rangle$ and their variations with the mass number of colliding nuclei, center-of-mass energy ($\sqrt{s_{NN}}$) and collision centrality are well supported by the available experimental data. Finally we give the predictions from our model for A-A collisions at the Large Hadron Collider (LHC) and Compressed Baryonic Matter (CBM) experiments. Our results indicate the existence of a possible universal production mechanism  for p-p, p-A and A-A collisions.   
\\
} 
\maketitle
\section{Introduction}
\label{intro}
\noindent
The ultimate goal of ultra-relativistic heavy-ion collision experiments is to test the predictions of quantum chromodynamics (QCD) because these can reveal the nature of hadronic interactions at extreme temperature and/or density and throw light on the role played by the quarks and gluons in  multiparticle production mechanism ~\cite{[singh]}. In order to verify the existence of a deconfined plasma created in nuclear collisions, we require accurate and precise knowledge about its subsequent hadronization. Hadronic multiplicities and their correlations can reveal enough information on the nature, composition and size of the fireball formed in the collider experiments~\cite{[kittel],[manjavidze]}. The evolution of the measured global observables in such experiments varies with the size and mass of the colliding nuclei and it emphasizes that rescatterings or multiple scatterings occur and  contribute to the evolution of the energy deposition in the fireball.

The search for some regularities and systematics present in the average multiplicity distributions observed in hadron-hadron (h-h), hadron-nucleus (h-A) and nucleus-nucleus (A-B) collisions has grown into a fascinating topic in strong interaction physics today because these properties hint at the underlying universal production mechanism for charged particles. The study of proton-proton ($p-p$) and $p-A$ collisions is particularly useful because it helps our understanding, e.g., in which manner $A-B$ collisions can differ from simple superposition of nucleon-nucleon collisions. However, our understanding of the subject is still quite poor partially because of the reason that in soft hadronic or nuclear collisions, the role of perturbative QCD is less clear as it falls close to the phase hadronization point where non-perturbative effects become clearly dominant. Several new experimental information on the multiparticle production have been reported in the recent past. Consequently a lot of efforts have also been put forward to understand and/or organise the experimental data by using various theoretical as well as phenomenological schemes~\cite{[amelin],[topor],[armesto],[drescher],[lin],[wang],[lokthin],[mitrovski],[sjostrand],[bopp],[sa],[deus],[albacete],[albacete1],[salgedo],[kharzeev],[kuiper],[choudhary],[capella],[el],[bzdak],[bialas],[humanic],[sarkisyan],[g]}. Some of these models explain multiplicities in nucleus-nucleus collisions mostly by using either quark-quark interactions or by considering basic nucleon-nucleon scatterings ~\cite{[bzdak],[bialas],[humanic],[sarkisyan]}. A compilation of various model calculations for charged hadron multiplicity that describe RHIC measurements at $\sqrt{s_{NN}}$ = 0.2 TeV  and for which predictions at $\sqrt{s_{NN}}$ = 2.76 TeV are available, has been shown in Ref.~\cite{[aamodt]}. Out of these  calculations, perturbative-QCD-inspired Monte carlo event generators, based on the HIJING model tuned upto 7 TeV pp data without jet quenching~\cite{[deng]}, or on the dual parton model~\cite{[w.bopp]}, or on the ultra relativistic quantum molecular dynamics (UrQMD)\cite{[mitrovski]}, show some consistency with the measurements. Models based on initial-state gluon density  saturation and gluon-gluon collisions have a range of predictions depending on the specific implementation~\cite{[albacete],[kharzeev],[levin],[e.levin],[n.armesto]} and exhibit a varying  level of agreement with the measurement. On the other hand, predictions of other models that successfully describe multiparticle production at RHIC energy usually differ at LHC energy by a factor of two~\cite{[aamodt]} which signifies a major discrepency. Thus, the mechanism for multiparticle production still needs a profound effort in order to clearly understand its dependence on the energy of the colliding nuclei. So we focus our efforts in formulating a phenomenological model which can describe multiplicity distributions consistently in the entire energy range accesible today at Relativistic Heavy Ion Collider (RHIC), Large Hadron Collider (LHC) and/or in future at Compressed Baryonic Matter (CBM) experiment on the FAIR (Facility of Antiproton and Ion Research) machine at GSI Darmstadt.

 The question as how one can describe the mechanics of such a relativistic many body problem in terms of simple, basic physical principles pose a major challenge in high energy heavy-ion physics today. Nucleus-nucleus collisions differ from nucleon-nucleon collisions in many respects. The presence of many constituent quarks participating in a chain of collisions makes the picture too complicated~\cite{[sakharov]}. In this paper, we describe nucleus-nucleus collisions in terms of effective number of constituent quarks participating in the collisions and the effective number of collisions suffered by each of them. In this model, we propose a gluon being exchanged between a quark of projectile or first nucleus and a quark belonging to target or other colliding nucleus. The resulting colour force is thus somewhat stretched between them and other constituent quarks because they try to restore the colour singlet behaviour. When two quarks separate, the colour force builds up a field between them and the energy in the colour field increases, the colour tubes thus formed finally break-up into new hadrons and/or quark-antiquark pairs. We essentially modify the phenomenological model earlier proposed by Singh et al ~\cite{[c.p.singh]}. We consider a multiple collision scheme in which a valence quark of the incident nucleon suffers one or more inelastic collisions with the quarks of target nucleons. In this process, quark loses energy and momentum, which results into creation of new hadrons. Further, we assume that all the quark collisions are independent and their effects are coherently superimposed. We find that charged particle multiplicities arising from $p-p$, $p-A$ and  $A-B$ collisions can be accounted in a consistent manner by adequately incorporating the number of participating quarks and mean number of quark-quark collisions. The rest of the paper is organised as follows. In section II, the detailed description of the model is presented, the results and discussions are summarized in section III and finally in section IV, main conclusions drawn from this study  are succinctly given.
 
\section{Description of Model}
\subsection{Total Mean Multiplicity of Charged Hadrons}
If we search a universal mechanism of charged particle production in the hadron-hadron, hadron-nucleus and nucleus-nucleus collisions, it must be driven by the available amount of energy required for the secondary production process, and it also depends on the mean number of participant quarks~\cite{[sakharov]}. The main ingredients of our model are taken from the paper by Singh et. al. ~\cite{[c.p.singh]}. Charged hadrons produced from $A-A$ collisions are assumed to have a somewhat unified production mechanism related to $p-p$ collisions at various energies. Feynman was first to point out that the multiplicity spectrum from proton-proton collisions becomes independent of the centre-of-mass energy ($\sqrt{s}$) as$\sqrt{s} \to \infty$~\cite{[feynman],[benecke],[hagedorn]}. This naturally implies that the total multiplicity after integration over rapidity involves $ln{\sqrt{s}}$ behaviour since $y_{max}=ln{\frac{\sqrt{s}}{m_{N}}}$, where $m_N$ is the nucleon-mass. Later on, it was realized that the gluons arising from gluon-bremsstrahlung processes give QCD - radiative corrections~\cite{[gunion]} and hence total multiplicity involves $ln^{2}\sqrt{s}$ behaviour~\cite{[b.bet],[jeon]}. Recently, it was noticed by PHOBOS collaboration from the $p-p$ and/or  $p-\bar{p}$ data that the central plateau height i.e. $(\frac{dn}{d\eta})_{\eta=0}$ grows as $ln^{2}\sqrt{s}$ which will give   $ln^{3}\sqrt{s}$ type behaviour to total multiplicity distribution~\cite{[b.bet],[jeon],[b.b]}.  

Based on the above experimental findings, recently Jeon and collaborators have shown that the total multiplicity obtained at RHIC can be bounded by a cubic logarithmic term in energy~\cite{[jeon]}. Therefore, we propose here a new parameterization involving a cubic logarithmic term  so that the entire $p-p$ experimental data ~\cite{[breakstone],[arnison],[ansorge],[alner],[abe],[whitmore],[grosse],[slattery],[thome]} starting from low energies (i.e. from 6.15 GeV) upto the highest LHC energy (i.e. 7 TeV)  can suitably be described:
\begin{equation}
 <n_{ch}> _{hp}=(a'+b' ln \sqrt{s_{a}}+c'ln^{2} \sqrt{s_{a}}+d'ln^{3} \sqrt{s_{a}})-\alpha.
\end{equation}
In Eq. (1), $\alpha$ is the leading particle effect and $\sqrt{s_{a}}$ is the available center-of-mass energy (i.e., $\sqrt{s_{a}}=\sqrt{s}-m_{B}-m_{T}$~\cite{[albini]}, where $m_{B}$ is the mass of projectile and $m_{T}$ the mass of the target nucleon, respectively), $a'$, $b'$, $c'$ and $d'$ are chosen constants. We take the values $a'=1.8$, $b'=0.37$, $c'=0.43$ and $d'=0.04$ as derived from the best fit to the data. The value of $\alpha$ is taken here as 0.85~\cite{[c.p.singh]}.
\begin{figure}
\resizebox{0.5\textwidth}{!}{%
  \includegraphics{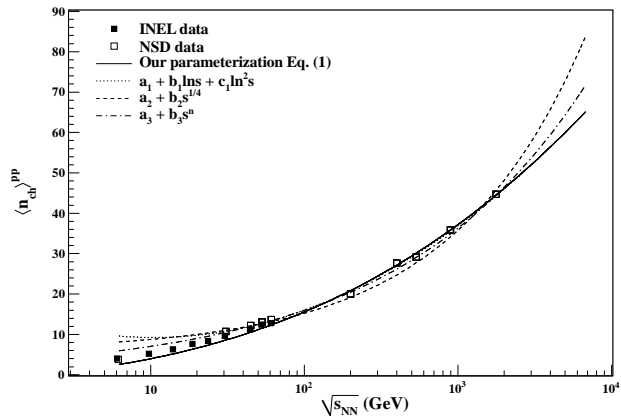}
}

\caption{Variations of total mean multiplicities of charged hadrons in $p-p$ collisions with $\sqrt{s_{NN}}$.}
\label{fig:1}       
\end{figure}

In Fig. (1), we present the inelastic (filled symbols) and non-single diffractive (NSD) data (open symbols) of charged hadron multiplicity in full phase space for $p-p$ collisions at various center-of-mass energies from different experiments e.g., ISR, UA5 and E735~\cite{[breakstone],[arnison],[ansorge],[alner],[abe],[whitmore],[grosse],[slattery],[thome]}. We use inelastic data at very low energies (filled symbols) because NSD data are not available for these energies and also the trend shows that the difference between inelastic and NSD data is very small at lower energies. Further, we also fit this data set with three other different functional forms in order to make a simultaneous comparison. The short-dashed line has the functional form as : $a_2+b_2s^{1/4}$ which is actually inspired by the Fermi-Landau model~\cite{[fermi],[c.-y.wong],[landau]}. It provides a resonable fit to the data at higher $\sqrt{s_{NN}}$ with $a_2 = 5.774$ and $b_2 = 0.948$~\cite{[grosse]}. However, since $a_2$ summarizes the leading particle effect also, its value  should not be much larger than two. The dotted line represents the functional form as : $a_1+b_1$ $ln s+c_1$ $ln^{2}s$~\cite{[albini]} and it fits the data well at higher $\sqrt{s_{NN}}$ but shows a large disagreement with the experimental data at lower center-of-mass energies with the values  $a_1 = 16.65$, $b_1 = -3.147$ and $c_1 = 0.334$, respectively. The dashed-dotted line represents the form $a_3+b_3s^{n}$~\cite{[albini]} and it provides a qualitative description of the data with $a_3=0, ~b_3=3.102$ and $n=0.178$ ~\cite{[grosse]}. The solid line represents our parametrization given by Eq. (1) and it offers a reasonable fit to the data starting from very low energy upto very high $\sqrt{s_{NN}}$ with the value of parameters as mentioned earlier in previous paragraph.

 We can extrapolate the validity of this parametrization for the produced charged particles in hadron-nucleus interactions by considering multiple collisions suffered by the quarks of hadrons in the nucleus. The number of constituent quarks which participate in hadron-nucleus (h-A) collisions share the total available  center-of-mass energy $\sqrt{s_{hA}}$ and thus the energy available for each interacting quark becomes $\sqrt{s_{hA}}/N_q^{hA}$~\cite{[c.p.singh]}, where $N_q^{hA}$ is the mean number of constituent quarks in h-A collisions. The total available squared center-of-mass energy  $s_{hA}$ in h-A collisions is related to $s_a$  as ${s_{hA}} = {\nu_q^{hA}s_a}$ ~\cite{[patashinskii],[chung]} with $\nu_{q}^{hA}$ as the mean number of inelastic collisions of quarks with target nucleus of atomic mass A. Within the framework of the Additive Quark Model~\cite{[anisovich],[a.bialas],[devidenko],[kobrinsky],[nyiri]}, the mean number of collisions in hadron-nucleus interactions is defined as the ability of constituent quarks in the projectile hadron to interact repeatedly inside a nucleus. The average number of constituent quarks ($N_{q}^{hA}$) interacting inelastically in h-A collisions is given by~\cite{[shyam]},
\begin{equation}
N_{q}^{hA} = \frac{\bar{\nu}(A)}{\bar{\nu}_{qA}},
\end{equation}
where, $\bar{\nu}(A)$ is the mean number of collisions of the incident hadron and $\bar{\nu}_{qA}$ represents the mean number of collisions of the wounded quarks inside the nucleus A and is defined by~\cite{[shyam]},
\begin{equation}
\bar{\nu}_{qA} = \frac{A\sigma_{qN}^{in}}{\sigma_{qA}^{in}},
\end{equation}
where $\sigma_{qN}^{in}$ and $\sigma_{qA}^{in}$ are the inelastic cross-sections for quark-nucleon (q-N) and quark-nucleus (q-A) interactions, respectively and A is the atomic mass of the target nucleus.

The $N_{q}^{hA}$~\cite{[c.p.singh],[shyam]} can be obtained from
\begin{equation}
N_{q}^{hA} = \frac{N_{c}\sigma_{qA}^{in}}{\sigma_{hA}^{in}}.
\end{equation}
Here $\sigma^{in}_{hA}$ is the inelastic hadron-nucleus cross section and $N_c$ is the number of valence quarks in the hadron h. The quark-nucleus inelastic cross-section ($\sigma_{qA}^{in}$) is determined from $\sigma_{qN}^{in}$ ($\approx$ $\frac{1}{3}\sigma_{NN}^{in} $) by using Glauber's approximation given as follows~\cite{[c.p.singh]}:
\begin{equation}
\sigma_{qA}^{in}=\int d^{2}b\left[1-\left(1-\sigma_{qN}^{in}D_{A}(b)\right)^{A}\right],
\end{equation}
where profile function $D_{A}(b)$ is related to the nuclear density, $\rho(b,z)$ by the relation~\cite{[c.p.singh]} :
\begin{equation}
D_{A}(b)=\int_{-\infty}^{\infty}\rho(b,z)dz,
\end{equation}
We have chosen the following form for the nuclear density as follows~\cite{[miller]}:
\begin{equation}
\rho(b,z)=\frac{\rho_{0}}{1+exp(\frac{\sqrt{b^{2}+z^{2}}-R}{a})},
\end{equation}
where all the notations have their usual meaning~\cite{[miller]}. We have taken the values of the constants i.e., $\rho_{0}, R$ and $a$ for different colliding nuclei separately from the Ref. ~\cite{[devries]} and b is transverse coordinate of the collision zone.

By using Eq. (4), one can easily see that the average number of constituent quark ($N_q$) is  1 for h-p collisions as the interactions occur between single constituent or dressed quark in accordance with the additive quark picture~\cite{[kobrinsky],[nyiri]} and other quarks are considered to be spectators. Thus the energy of the initial thermalized state which is responsible for the production of secondary particles is assumed to arise mainly from interacting single quark pair. The spectator quarks which are not the part of thermalized volume at the moment of collision, do not thus participate in secondary particle production process. As a result, the leading particles~\cite{[basile],[m.basile]} resulting from the spectator quarks carry away a significant part of the energy~\cite{[sakharov]}. 
Finally, the expression for average charged hadron multiplicity in h-A collisions is~\cite{[c.p.singh]}:

\begin{eqnarray}
 <n_{ch}> _{hA}=N_{q}^{hA}&\Biggl[&a'+b' ln \left(\frac{\sqrt{s_{hA}}}{N_{q}^{hA}}\right)+c'ln^{2}\left(\frac{\sqrt{s_{hA}}}{N_{q}^{hA}}\right) \nonumber \\
&+&d'ln^{3}\left(\frac{\sqrt{s_{hA}}}{N_{q}^{hA}}\right)\Biggr] - \alpha.
\end{eqnarray}

The generalization of the above picture for the case of nucleus-nucleus collisions  can be achieved as follows~\cite{[c.p.singh]}:
\begin{eqnarray}
<n_{ch}> _{AB}=N_{q}^{AB}&\Biggl[&a'+b' ln \left(\frac{\sqrt{s_{AB}}}{N_{q}^{AB}}\right)+c'ln^{2}\left(\frac{\sqrt{s_{AB}}}{N_{q}^{AB}}\right) \nonumber \\
&+&d'ln^{3}\left(\frac{\sqrt{s_{AB}}}{N_{q}^{AB}}\right) \Biggr],
\end{eqnarray}

where $\sqrt{s_{AB}} = A \sqrt{\nu_q^{AB}s_a}$ and the mean number of inelastic quark collisions $\nu_q^{AB}$ is:
\begin{equation}
\nu_{q}^{AB}=\nu_{qA}\nu_{qB}=\frac{A\sigma_{qN}^{in}}{{\sigma_{qA}^{in}}}.\frac{B\sigma_{qN}^{in}}{{\sigma_{qB}^{in}}}.
\end{equation}
Furthermore, mean number of participating quarks $N^{AB}_{q}$ in A-B collisions can be calculated by generalizing Eq. (4):
\begin{equation}
N^{AB}_{q}=\frac{1}{2}\left[\frac{N_{B}\sigma_{qA}^{in}}{{\sigma_{AB}^{in}}}+\frac{N_{A}\sigma_{qB}^{in}}{{\sigma_{AB}^{in}}}\right],
\end{equation}
where $\sigma_{AB}^{in}$ is the inelastic cross-section for nucleus-nucleus (A-B) collision and can be obtained as~\cite{[c.p.singh],[huang]}:
\begin{equation}
\sigma_{AB}^{in}= \pi r^{2}\left[A^{1/3}+B^{1/3}-\frac{c}{A^{1/3}+B^{1/3}}\right]^2,
\end{equation}
where c is a constant and has a value 4.45 for nucleus-nucleus collisions. 

The parametrization in Eq. (9) thus relates nucleus-nucleus collisions to hadron-nucleus and hadron-proton collisions and the values of the parameters $a', b', c'$, and $d'$ remain unaltered which shows its universality for all these processess.

In creating quark gluon plasma (QGP), greater emphasis is laid on the central or head-on collisions of two nuclei. The mean multiplicity  in central collisions can straight forwardly be generalized from Eq. (9) as :
\begin{eqnarray}
<n_{ch}>^{central}_{AB}=A&\Biggl[&a'+b'ln(\nu_{q}^{AB}s_{a})^{1/2}+c'ln^{2}(\nu_{q}^{AB}s_{a})^{1/2}\nonumber \\
&+&d'ln^{3}(\nu_{q}^{AB}s_{a})^{1/2}\Biggr].
\end{eqnarray}
\subsection{Pseudo-rapidity Distributions}
 The pseudo-rapidity distribution of charged particles is another important quantity in the studies of particle production mechanism from high energy $h-h$ and $A-B$ collisions, which, however, is not yet understood properly. It has been pointed out that $(dn_{ch}/d\eta)$ can be used to get the information on the temperature ($T$) as well as energy density ($\rho$) of the QGP ~\cite{[x.n.wang],[sousa],[eskola]}. To calculate the pseudo-rapidity density of charged hadrons, we first fit the experimental data of $(dn_{ch}/d\eta)^{pp}_{\eta=0}$ for collision-energy ranging from  a low energy to very high energy. One should use the  parameterization upto squared logarithmic term in accordance with  Ref.~\cite{[jeon]}.
Hence,  using a parameterization for central rapidity density as:

\begin{figure}
\resizebox{0.5\textwidth}{!}{%
  \includegraphics{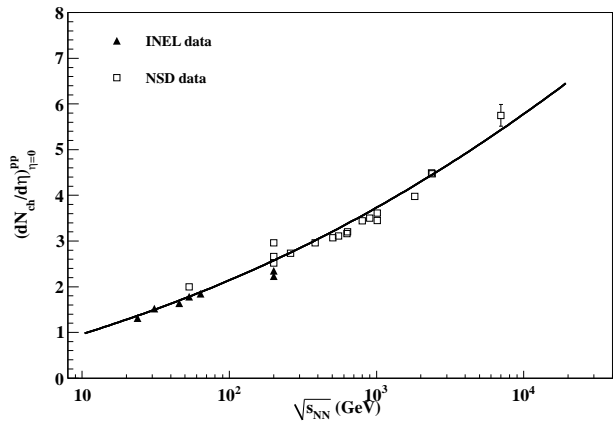}
}

\caption{Variations of pseudorapidity density at mid-rapidity for $p-p$ collisions with $\sqrt{s_{NN}}$. Filled symbols are data from inelastic $p-p$ events and open symbols are experimental data for NSD events~\cite{[arnison],[ansorge],[alner],[abe],[thome],[sun],[noucier],[i.abelev],[k.aamodt],[khachatryan]}. Solid line is the outcome of our parameterization.}
\label{fig:4}       
\end{figure}
\begin{eqnarray}
 <(dn_{ch}/d\eta)^{pp}_{\eta=0}> &=&(a_{1}^{'}+b_{1}^{'} ln \sqrt{s_{a}}+c_{1}^{'}ln^{2} \sqrt{s_{a}})-{\alpha_1}',\nonumber \\
\end{eqnarray}
we obtain the values of the parameters $a_{1}' = 1.24$, $b_{1}' = 0.18$, and $c_{1}' = 0.044$ from the reasonable fit to the p-p data. We notice that the value of ${\alpha_1}'$ is very small but not negligible.  We also notice that the numerical integration of pseudo-rapidity distribution in A-A collisions, using the functional form of Eq. (14), exactly tallies with the total multiplicity of charged hadrons in central A-A collision as obtained from Eq. (13). 

In Fig. 2, we present the inelastic (filled symbols) and non-single diffractive (NSD) data (open symbols) of $\frac{dn_{ch}}{d\eta}$ at mid-rapidity for $p-p$ collisions at various center of mass energies from different experiments e.g., ISR, UA5, E735, RHIC and LHC ~\cite{[arnison],[ansorge],[alner],[abe],[thome],[sun],[noucier],[i.abelev],[k.aamodt],[khachatryan]}. Further, solid line represents our parameterization fit with the values of the parameters given above.

Earlier many authors have attempted to calculate the pseudo-rapidity density of charged hadrons in a two component model of partons fragmentation~\cite{[trainor],[a.trainor]}. Its physical interpretation is based on a simple model of hadron production: longitudinal projectile nucleon dissociation (soft) and transverse large-angle scattered parton fragmentation (hard). In A-A collisions hadron production from the two process scales as N$_{part}$/2 (number of nucleon participant pairs) and N$_{bin}$ (number of binary N-N collisions), respectively~\cite{[d.kharzeev]}. However, this assumption which is based on a nucleon-nucleon collisional Glauber model is crude and it looks unrealistic to relate participating nucleons and nucleon-nucleon binary collisions to soft and hard components at the partonic level. Here we modify the two component model of pseudo-rapidity distributions in A-A collisions in the wounded quark scenario and assume that the hard component which  basically arises due to multiple parton interactions~\cite{[t.a.trainor]} scales with the number of quark-quark collisions (i.e. $N_{q}^{AB}\nu_{q}^{AB}$) and soft component scales with the number of participating quarks (i.e. $N_{q}^{AB}$). Thus, the expression for $(dn_{ch}/d\eta)^{AB}_{\eta=0}$ in A-B collisions can be parameterized in terms of p-p rapidity density as follows  :
\begin{equation}
\left(\frac{dn_{ch}}{d\eta}\right)^{AB}_{\eta=0}=\left(\frac{dn_{ch}}{d\eta}\right)^{pp}_{\eta=0}\left[\left(1-x\right)N_{q}^{AB}+ x N_{q}^{AB}\nu_{q}^{AB}\right],
\end{equation}
where $x$ quantifies the relative contributions of two components arising from hard and soft processes. The fraction $x$ corresponds to the hard processes and the remaining fraction ($1-x$) describes the contributions from the soft processes. We take the value of $x$ varying from  $0.1-0.125$ at different  $\sqrt s_{NN}$. Thus, the values of $x$  in the wounded quark model are found in close agreement with the values of $x$ used in Refs.~\cite{[d.kharzeev],[ande.levin],[s.y.li],[phobos]}. 

We further extend the model in order to incorporate $\eta$ dependence in central A-B collisions by using the functional form as given in Ref.~\cite{[b.alver]}: 

\begin{equation}
(\frac{dn_{ch}}{d{\eta}})^{AB} = 2(\frac{dn_{ch}}{d\eta})^{AB}_{\eta = 0} \frac{\sqrt{1- \frac{1}{(\beta cosh {\eta})^2}}}{\gamma + exp (\eta^2/2{\sigma}^2)},
\end{equation}

where $\beta, \gamma$ and $\sigma$ are fitting parameters and $({dn_{ch}}/{d\eta})^{AB}_{\eta = 0}$ is the central pseudo-rapidity density in  A-B collisions obtained from Eq. (15). In Eq. (16), the factor $\sqrt{1- \frac{1}{(\beta cosh {\eta})^2}}$  is responsible for the dip in the mid rapidity region of the distribution. Parameter $\sigma$ is responsible for the width of the pseudo-rapidity distribution. 

One can question our intentions in explaining the total multiplicity $\langle n_{ch} \rangle$, ${dn_{ch}}/{d{\eta}}$ at $\eta=0$, and ${dn_{ch}}/{d{\eta}}$ for the nucleus-nucleus collisions where we know these quantities in $p-p$ or $h-p$ collisions. We have to use additional physics input in the whole exercise eg. the physics from two component model and giving a width for $\eta$-dependence for understanding the complete spectrum of the data. The values of fitting parameters used in Eq. (16) for pseudo-rapidity distributions for various colliding nuclei at different $\sqrt {s_{NN}}$ are shown in Table 1. 

\section{Results and Discussions}

\begin{figure}
\resizebox{0.5\textwidth}{!}{%
  \includegraphics{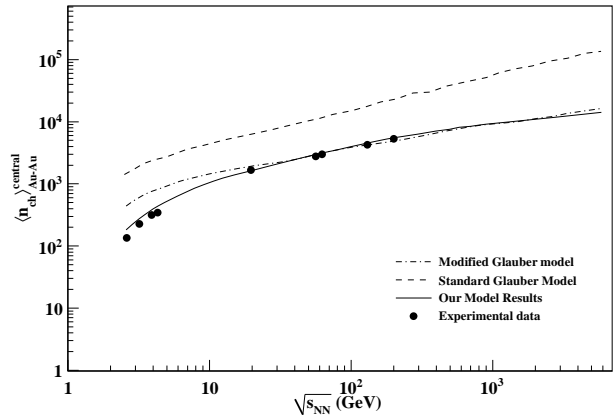}
}
\caption{Variations of total mean multiplicities of charged hadrons in central $Au-Au$ collisions with $\sqrt{s_{NN}}$. Experimental data points are taken from Ref.~\cite{[b.alver],[adler],[b.b.back],[klay]}.}
\label{fig:2}       
\end{figure}

\begin{figure}
\resizebox{0.5\textwidth}{!}{%
  \includegraphics{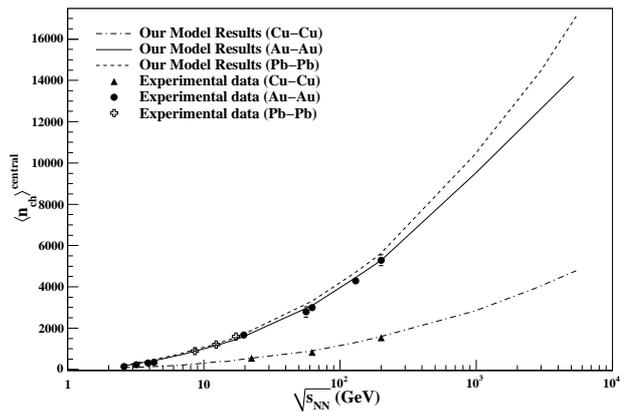}
}

\caption{Variations of total mean multiplicities of charged hadrons in central collisions for different colliding nuclei with $\sqrt{s_{NN}}$.}
\label{fig:3}       
\end{figure}

Fig. 3, shows the variation of mean multiplicity of charged hadrons produced in central $Au-Au$ collision with respect to $\sqrt{s_{NN}}$. We also compare our model results (solid line) with the standard Glauber model (dashed line) and the modified Glauber model results (dash-dotted line) ~\cite{[feofilov]} with the experimental data of AGS and RHIC~\cite{[b.alver],[adler],[b.b.back],[klay]}. We would like to mention here the main difference between modified Glauber model and the standard Glauber calculation. In standard Glauber model, a participating nucleon can have one or more number of collision at the same $\sqrt{s_{NN}}$ and using a constant value for $\sigma_{inel}^{pp} (\sqrt{s})$. However, in modified Glauber model, a participating nucleon loses a fraction of its momentum after the first collision and participate in subsequent collisions with somewhat lower energy and also a lower value of $\sigma_{inel}^{pp} (\sqrt s)$. This modification significantly suppresses the number of collisions in comparison to that obtained in the standard Glauber model. We find that the results obtained in our model give a better description to the experimental data in comparison to the standard as well as modified Glauber model predictions. This clearly shows the significance of the role played by quark degrees of freedom as used by us in our picture.
\begin{table*}
\begin{center}
\caption{The values of fitting parameters used in Eq. (16) for pseudo-rapidity  distribution for various colliding nuclei at different $\sqrt{s_{NN}}$. }
\begin{tabular}{|l|l|l|l|l|l|l|l|l|l|l|l|l}
\hline
{$\sqrt{s_{NN}}$ (GeV)} & \multicolumn{3}{l|}{{\qquad Cu-Cu }}&\multicolumn{3}{l|}{{\qquad Au-Au }} &\multicolumn{3}{l|}{{\qquad Pb-Pb}}  \\
\cline{2-10}
 &{$\beta$}&{$\gamma$ }&{$\sigma$ } &{$\beta$}&{$\gamma$ }&{$\sigma$ } &{$\beta$}&{$\gamma$ }&{$\sigma$ }\\
\hline\hline

  $19.6$   & --      & --     & --    &1.79   & 0.46     &1.55       & --    & --   &--  \\
$22.4$     & 1.66  & 0.54 &1.64 & --      &  --        & --          & --    &--    &--    \\
$62.4$     & 1.75  & 0.86 &2.00 &1.69   & 0.85     &2.00       &--     &--    &--      \\
$130$      &  --     &  --    &  --   &1.76   & 0.95     &2.18       &--     & --   &--       \\
$200$      &  1.80 &  0.89&2.47 & 1.79  & 0.99     &2.29       &--     & --   &--         \\
$2760$     &--       & --     &--     & --      & --         & --          &1.82 &0.80& 3.28         \\ \hline

\end{tabular}
\end{center}
\end{table*}

In Fig. 4, we show the variation of mean multiplicity calculated from our model for $Au-Au$, $Pb-Pb$ and $Cu-Cu$ central collisions with respect to $\sqrt{s_{NN}}$. We also compare our model results with the experimantal data of RHIC and SPS experiments~\cite{[b.alver],[adler],[b.b.back],[klay],[afanasiev]}. One can see from Fig. 4, that the model results give an excellent fit to the separate experimental data points for nucleus of varying sizes, eg., $Cu-Cu$,  $Au-Au$ and $Pb-Pb$, respectively. This endorses our view point in considering the superposition of basic quark-quark collisions used in describing the nucleus-nucleus collision experiments with nuclei of different sizes.

In Table 2, we have shown the mean multiplicity of charged hadrons in minimum-bias as well as in central $d-Au$ collisions at $\sqrt{s_{NN}}=200$ GeV as obtained in our model. We have also shown the experimental results~\cite{[b.b.back et]} for this collision. The good agreement between the experimental and model results again shows that it can be used for asymmetric colliding nuclei as well.

\begin{figure}
\resizebox{0.5\textwidth}{!}{
  \includegraphics{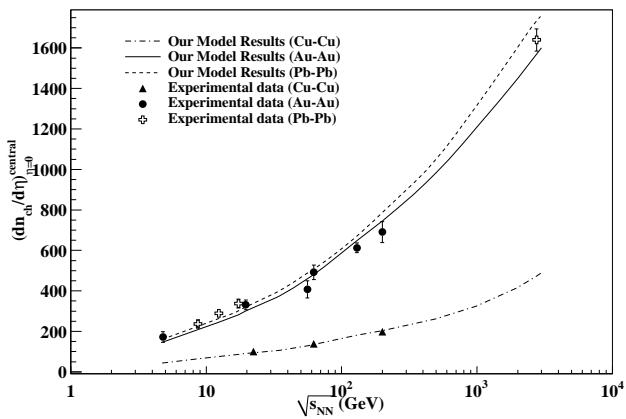}
}

\caption{Variations of pseudorapidity density at mid-rapidity for $A-A$ collision with $\sqrt{s_{NN}}$. Symbols are data from various heavy ion collision experiments ~\cite{[b.alver],[adler],[b.b.back],[klay],[afanasiev],[k.aamodt et],[s.chatrchyan]}.}
\label{fig:5}       
\end{figure}

In Fig. 5, we plot the variations of $(dn_{ch}/d\eta)_{\eta=0}$ obtained in our model for $Au-Au$, $Pb-Pb$ and $Cu-Cu$ central collisions with respect to $\sqrt{s_{NN}}$. We also compare our results with the experimental data of RHIC, SPS and LHC experiments ~\cite{[b.alver],[adler],[b.b.back],[klay],[afanasiev],[k.aamodt et],[s.chatrchyan]}. Our model results compare well with the experimental data in the entire energy range. As we didn't take any effect of final-state interactions in our model, it hints at the fact that the hadron multiplicity in $A-B$ collisions is mainly driven by the initial parton production and hence the effect of final-state interaction should be negligibly small as suggested in some earlier calculations ~\cite{[j.eskola],[w.lin]}.
\begin{table}

\begin{center}
\caption{Total mean multiplicity of charged hadrons produced in $d-Au$ collision at $\sqrt{s_{NN}}=200$ GeV.}
\resizebox{9cm}{!}{
\begin{tabular}{cccc} \hline \hline 
                     &Colliding Nuclei     & $\langle n_{ch}\rangle^{min. bias}$        & $\langle n_{ch}\rangle^{central}$                \\ 
         
\hline 
Our Model            & $d-Au$               & $82$                             & $162$      \\
PHOBOS ~\cite{[b.b.back et]}         & $d-Au$               & $87_{-6}^{+7}$                 & $167_{-11}^{+14}$    \\ \hline
\end{tabular}
}
\end{center}
\end{table} 

\begin{figure}
\resizebox{0.5\textwidth}{!}{
  \includegraphics{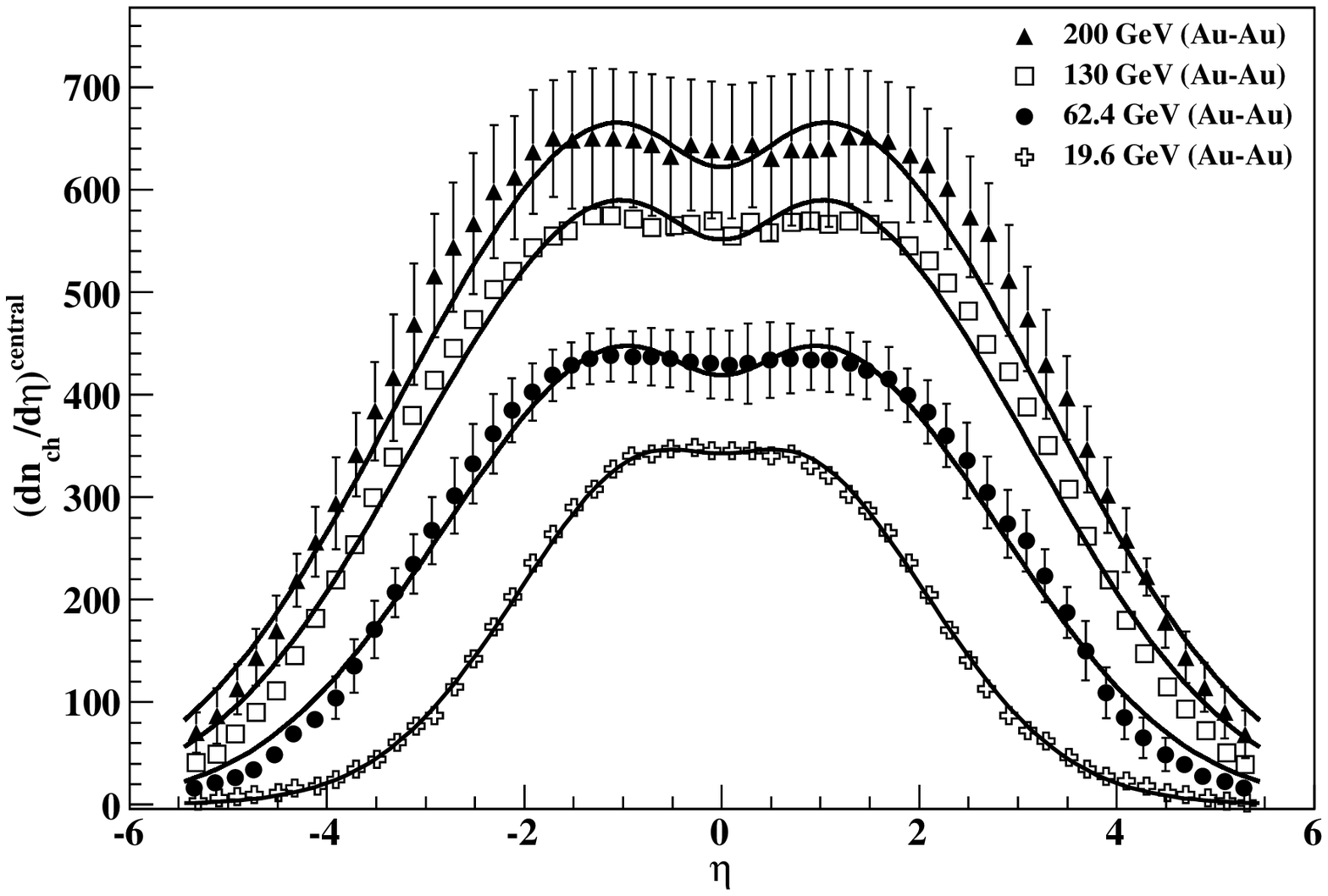}
}

\caption{Variations of pseudo-rapidity density in the 6$\%$} most-central Au-Au collisions at RHIC energies. Data are taken from Ref.~\cite{[b.b.backet.al.]}. Solid line curves represent our model results.
\label{fig:6}       
\end{figure}

\begin{figure}
\resizebox{0.5\textwidth}{!}{
  \includegraphics{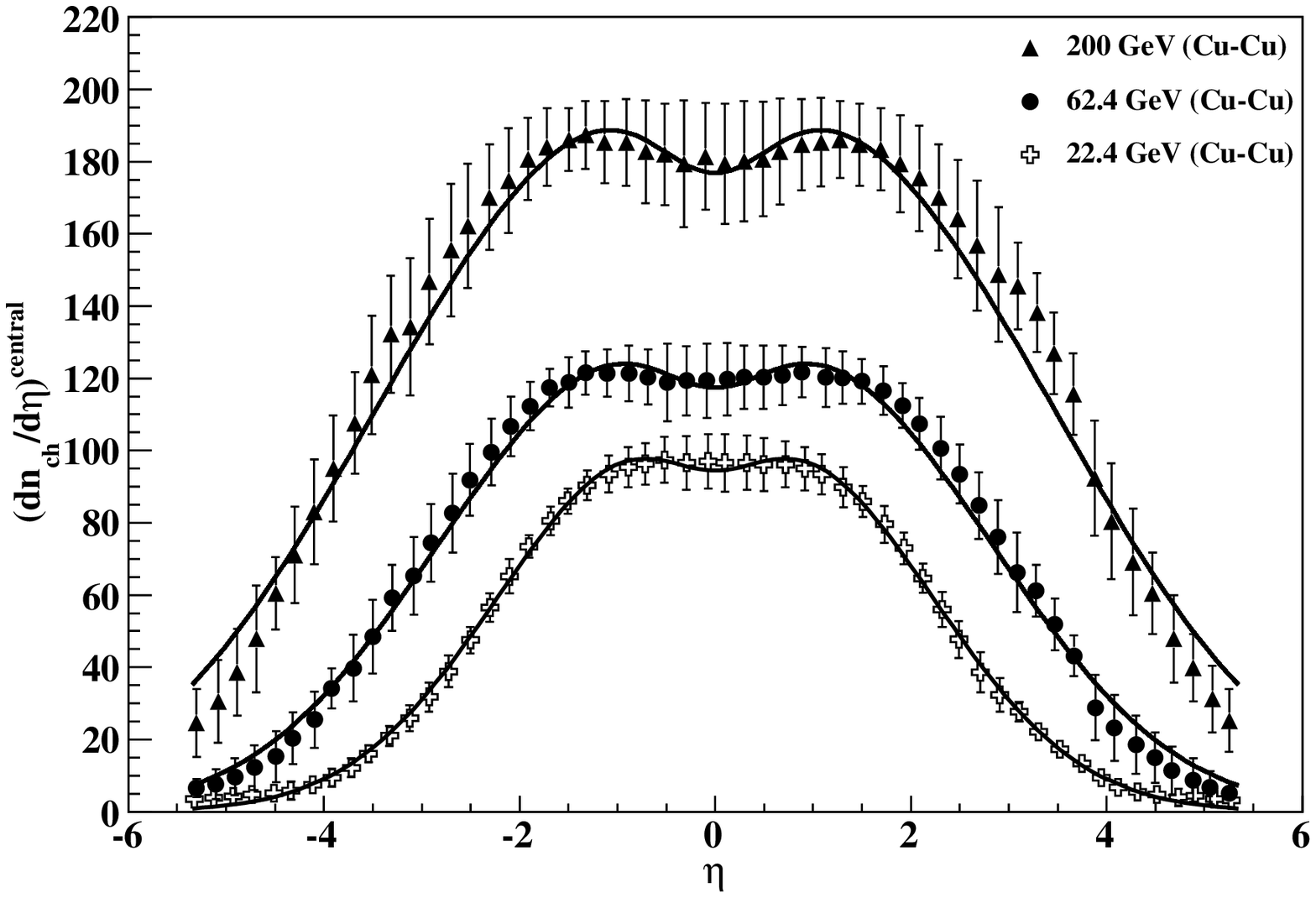}
}
\caption{Variations of pseudo-rapidity density in the 6$\%$} most-central Cu-Cu collisions at RHIC energies. Data are taken from Ref.~\cite{[veres]}. Solid line curves represent our model results.
\label{fig:7}       
\end{figure}
In Fig. 6, we have shown the variations of pseudo-rapidity densities obtained in our model for central Au-Au collisions with respect to $\eta$ and are compared with the experimental data of RHIC~\cite{[b.b.backet.al.]} at four energies ($\sqrt{s_{NN}}$ = 19.6, 62.4,130 and 200 GeV). Similarly, in Fig. 7, we have shown the variations of pseudo-rapidity densities for central Cu-Cu collisions and is compared with the experimental data~\cite{[veres]} of RHIC at three energies ($\sqrt{s_{NN}}$ = 22.4, 62.4 and 200 GeV). In Fig. 8, we have shown the variation of pseudo-rapidity densities obtained in our model for central Pb-Pb collisions at LHC energy and is compared with the preliminary ALICE data~\cite{[toia]}.  At LHC energies, the effect of the factor ($\sqrt{1-\frac{1}{(\beta cosh\eta)^2}}$) in Eq. (16) for a given particle mass tends to be smaller due to the higher  average transverse momenta~\cite{[wolschin]}, but the dip in the preliminary ALICE data is clearly more pronounced~\cite{[toia]} than at RHIC~\cite{[david]}. The sizable dip at midrapidity seen in the preliminary ALICE data at 2.76 TeV is easily achieved in our results also.  The good agreement with the experimental data clearly shows that the model is capable of providing a clear $\eta$ dependence and it can be used to reproduce the pseudo-rapidity data at various energies with different set of colliding nuclei although at the cost of some extra parameters as used in Eq. (16). Furthermore, it is quite obvious from the agreement with the experimental data that integral of the pseudo-rapidity density over full pseudo-rapidity range in our model should match with the total multiplicity in central nucleus-nucleus (A-A) collisions.  

\begin{figure}
\resizebox{0.5\textwidth}{!}{
  \includegraphics{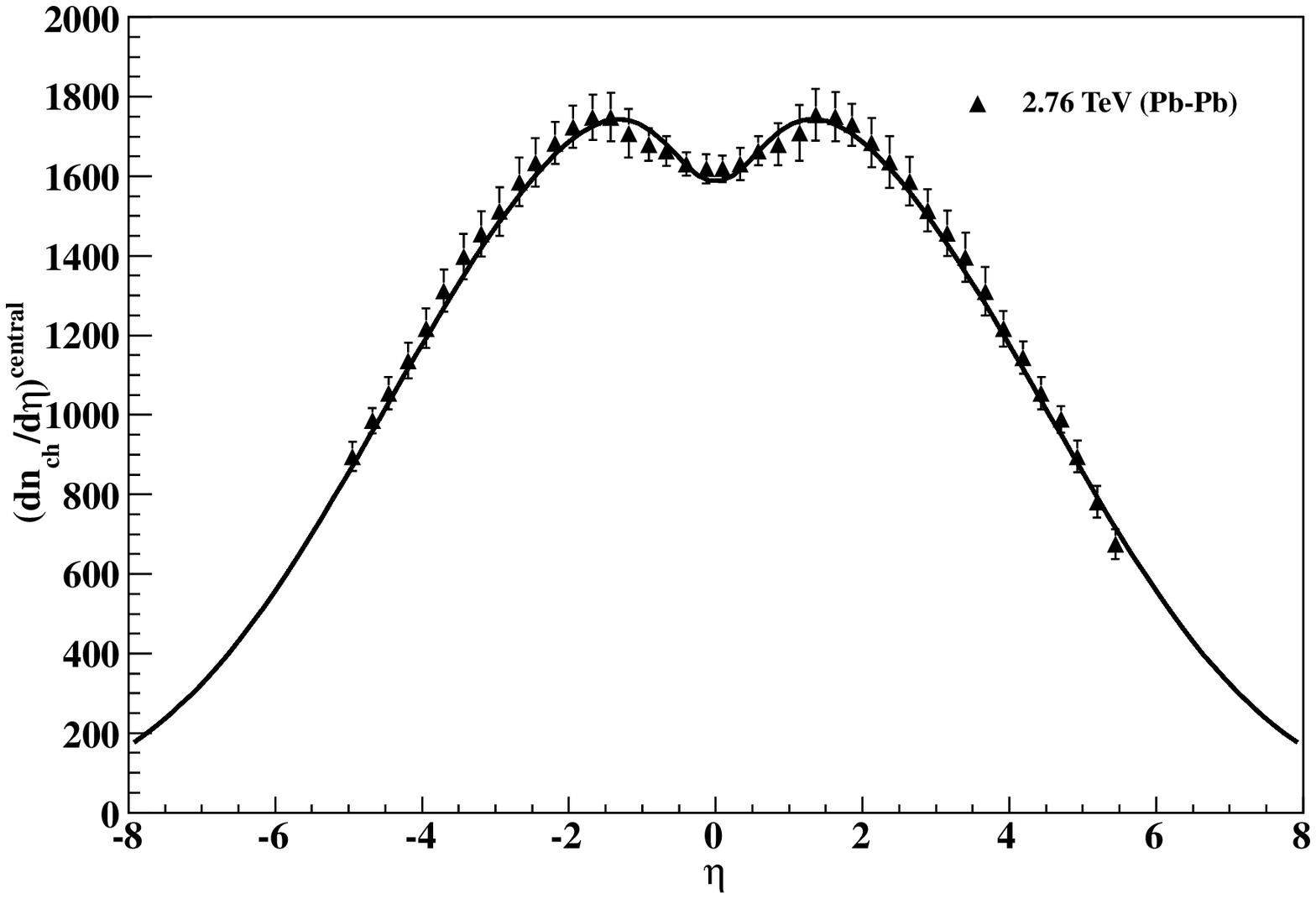}
}
\caption{Variations of pseudo-rapidity density in the 5$\%$} most-central Pb-Pb collisions at LHC energy (2.76 TeV). Data are taken from Ref.~\cite{[toia]}. Solid line curve represent our model results.
\label{fig:8}      
\end{figure}

\begin{figure}
\resizebox{0.5\textwidth}{!}{
\includegraphics{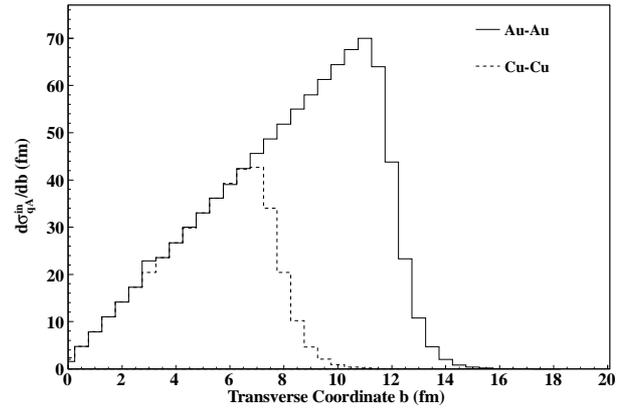}
}
\caption{Variation of $d\sigma^{in}_{qA}/db$ in our model as a function of transverse coordinate ($b$) of collision zone formed in Au-Au and Cu-Cu collisions.}
\label{fig:9}      
\end{figure}

\begin{figure}
\resizebox{0.5\textwidth}{!}{
\includegraphics{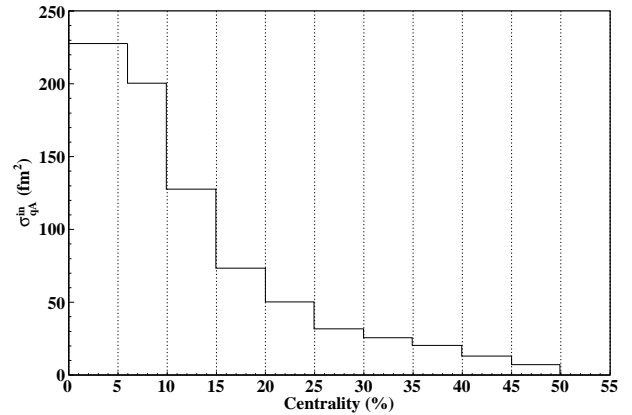}
}
\caption{Variation of quark-nucleus inelastic cross-section ($\sigma^{in}_{qA}$ ) in our model as a function of centrality for Au-Au collisions.}
\label{fig:10}      
\end{figure}
\begin{table*}
\begin{center}
\caption{The total charged hadron multiplicities as a function of centrality in Au-Au Collisions at three different RHIC energies. Exerimental data are taken from Ref.~\cite{[b.b.backet.al.]}}
\begin{tabular}{|l|l|l|l|l|l|l|l|l|l|l|l|l}
\hline
{Centrality Bin} & \multicolumn{2}{l|}{{$\sqrt{s_{NN}}=200$ GeV}}&\multicolumn{2}{l|}{{$\sqrt{s_{NN}}=130$ GeV}} &\multicolumn{2}{l|}{{$\sqrt{s_{NN}}=62.4$ GeV}}  \\
\cline{2-7}
 &{Model}&{Experimental }&{Model }&{Experimental }&{Model }&{Experimental }\\
\hline\hline

  ~$0-6 \%$  & 5277     &5095$\pm$255   & 4449       & 4195$\pm$210     &3098     &2881$\pm$143           \\
 ~$6-10 \%$  & 4177     &4341$\pm$245   & 3480       & 3649$\pm$182       &2354     &2489 $\pm$124             \\
$10-15 \%$   & 3601     &3763$\pm$188   & 3072       & 3090$\pm$155       &2196     &2120$\pm$106           \\
$15-20 \%$   & 2939     &3153$\pm$158   & 2554       & 2586$\pm$129       &1886     &1777$\pm$88                \\
$20-25 \%$   & 2492     &2645$\pm$132   & 2189       & 2164$\pm$108       &1673     &1485$\pm$74             \\
$25-30 \%$   & 2003     &2184$\pm$109   & 1779       & 1793$\pm$90        &1392     &1236$\pm$61              \\
$30-35 \%$   & 1790     &1819$\pm$91    & 1596       & 1502$\pm$75        &1262     &1027$\pm$51              \\
$35-40 \%$   & 1591     &1486$\pm$74    &1428        & 1222$\pm$61        &1104     &840$\pm$42            \\
$40-45 \%$   & 1230     &1204$\pm$60    & 1111       & 975$\pm$49         &902      &679$\pm$33               \\ 
$45-50 \%$   & 881      &951$\pm$48     & 801        & 782$\pm$39        &662      &532$\pm$26 \\ \hline
\end{tabular}
\end{center}
\end{table*}

\begin{table*}
\begin{center}
\caption{The total charged hadron multiplicities as a function of centrality in Cu-Cu Collisions at three different RHIC energies. Exerimental data are taken from Ref.~\cite{[b.b.backet.al.]}}
\begin{tabular}{|l|l|l|l|l|l|l|l|l|l|l|l|l}
\hline
{Centrality Bin} & \multicolumn{2}{l|}{{$\sqrt{s_{NN}}=200$ GeV}}&\multicolumn{2}{l|}{{$\sqrt{s_{NN}}=62.4$ GeV}} &\multicolumn{2}{l|}{{$\sqrt{s_{NN}}=22.4$ GeV}}  \\
\cline{2-7}
 &{Model}&{Experimental }&{Model }&{Experimental }&{Model }&{Experimental }\\
\hline\hline

  ~$0-6 \%$  & 1532       &1474$\pm$69  & 884         &807$\pm$35       &483     &508$\pm$22    \\
 ~$6-10 \%$  & 1203       &1262$\pm$59  & 654         &721$\pm$32       &395     &431$\pm$19             \\
$10-15 \%$   & 1003       &1084$\pm$51  & 609         &635$\pm$27       &355     &375$\pm$18         \\
$15-20 \%$   & 850        &917$\pm$43   & 544         &541$\pm$24       &339     &320$\pm$15         \\
$20-25 \%$   & 708        &771$\pm$38   & 470         &460$\pm$21       &308     &273$\pm$14           \\
$25-30 \%$   & 600        &645$\pm$32   & 411         &386$\pm$17       &279     &230$\pm$13            \\
$30-35 \%$   & 504        &538$\pm$27   & 354         &323$\pm$15       &247     &194$\pm$12              \\
$35-40 \%$   & 400        &444$\pm$23   & 288         &270$\pm$13       &207     &162$\pm$12            \\
$40-45 \%$   & 325        &364$\pm$19   & 239         &223$\pm$11       &175     &135$\pm$11               \\ 
$45-50 \%$   & 270        &293$\pm$15   & 201         &183$\pm$9        &150     &112$\pm$11 \\ \hline

\end{tabular}
\end{center}
\end{table*}
Further, we have also attempted to explore the applicability of the model to exhibit  centrality dependence in nucleus-nucleus (A-B) collisions by varying the transverse coordinate ($b$) as used in Eq. (5). Transverse coordinate ($b$) determines the size of the collision zone formed during the collision, which is directly related to the particle multiplicity or the energy produced in each inelastic collision via inelastic quark-nucleus inelastic cross-section ($\sigma^{in}_{qA}$). The quark-nucleus inelastic cross-section in the collisions of two nuclei A and B, that is the integral of the distributions $d\sigma^{in}_{qA}/db$, is a basic quantity for determining the observed multiplicity in final state as it directly affects the mean number of participating quarks ($N_{q}^{AB}$) and the mean number of inelastic quark collisions ($\nu_{q}^{AB}$). In Fig. 9, we have shown the variations of $d\sigma^{in}_{qA}/db$ in our model as a function of $b$ of collision zone for Au-Au and Cu-Cu collisions. Furthermore, we have shown the variation of quark-nucleus inelastic cross-section ($\sigma^{in}_{qA}$ ) in the model as a function of centrality for Au-Au collisions in Fig. 10 to illustrate how the model works for the determination of the centrality in nucleus-nucleus collisions. Finally, the model results are organized in tabular form for ten centrality classes and are compared accordingly with the RHIC multiplicity data~\cite{[b.b.backet.al.]}. For suitable comparison of model results for most central ($0-6\%$) case in Au-Au and Cu-Cu collisions, we have taken the average of the RHIC multiplicitiy data~\cite{[b.b.backet.al.]} of first two centrality bins ($0-3\% ~ \& ~ 3-6\%$).  In Table 3, we have depicted a tabular representation of our model results for the total charged hadron multiplicities as a function of the centrality in Au-Au collisions at three RHIC energies ($\sqrt{s_{NN}}$ = 62.4, 130 and 200 GeV). We have also made a comparison with the experimental data~\cite{[b.b.backet.al.]} at the respective RHIC energies. Model results are in reasonable agreement with the RHIC data within experimental errors for Au-Au collisions. Similarly in Table 4, we have shown our model results for the total charged hadron multiplicities as a function of the centrality in Cu-Cu collisions at three RHIC energies ($\sqrt{s_{NN}}$ = 22.4, 62.4 and 200 GeV) and compared with the experimental data~\cite{[b.b.backet.al.]} at the respective RHIC energies. Our results again agree reasonably well with the RHIC data within experimental errors.    

In Table 5, we give our predictions for the mean multiplicity of charged hadrons, produced in minimum-bias as well as in central events at LHC and CBM energies. Specifically, we calculate the charged hadron multiplicity in $Pb-Pb$ collisions at $\sqrt{s_{NN}}= 2.76$ TeV for LHC and the mean multiplicity of charged hadrons in $Au-Au$ collisions at $E_{lab}= 8$ AGeV and $35$ AGeV for future CBM experiment. We also compare our results with the other model predictions like hadron string dynamics (HSD) model for CBM experiment~\cite{[cbm],[cassing]} and model of W. Busza for LHC experiment ~\cite{[armesto],[busza]}, respectively. We find that the predictions from our calculations are reasonably supported by their findings.

\section{Conclusions}
Within the framework of constituent quarks, we have given a parametrization which interrelates the multiplicity and mid-rapidity density distributions in p-p, p-A and A-A collisions from a few GeV upto the highest LHC energies. It involves the interactions of the constituent quarks of the colliding objects or beam and target particles or nuclei and the mean number of collisions suffered by them. We assume that the production of secondary particles is proportional to the fraction of the available energy for the participant quarks. This picture describes consistently the `soft' hadron production in $p-p$, $p-\bar{p}$, $p-A$ and $A-A$ collisions. In heavy-ion collisions, one participating quark can have larger number of interactions due to the large size of the nucleus as well as due to a large travel path to be travelled inside the nucleus and hence, we get larger energy  available for the secondary particle production~\cite{[sakharov]}. We find that our parametrization gives an excellent fit to the $p-p$ data in the entire energy range. We also find that a suitable extension of this description works very well for the nucleus-nucleus ($A-A$) collisions in the entire energy range. We have compared our model results with the results obtained from the standard as well as modified Glauber model to demonstrate the difference between the role played by nucleon-nucleon interactions vis-a-vis quark-quark interactions. We have also shown the charged particle multiplicity in central as well as in minimum-bias $d-Au$ collisions at $\sqrt{s_{NN}}=200$ GeV and we find the results compare quite well with the experimental results. This exercise clearly shows that our model also works quite well for asymmetric colliding nuclei. We have also proposed an extension of the two-component model based on our parameterization and the results for rapidity distribution obtained by us again work well. We have given our predictions for LHC and CBM machines as well and compare with other model predictions~\cite{[armesto],[busza]} existing in the literature. 

 In conclusion, we have attempted to point out a mechanism for `soft' hadron production in hadron-hadron, hadron-nucleus and nucleus-nucleus collisions. Due to non-perturbative nature, QCD is not expected to work for these cases. So lacking a workable theory, we have relied on the phenomenology and all the salient features of the experimental data are explained by our parameterization. We hope that the work done here will highlight the multiparticle production mechanism present in nucleus-nucleus collisions in relation to p-p interactions which can be of use in detecting any deviation observed in the data from the predictions of our model and thus a hint for QGP formation~\cite{[singh]} can be obtained in ultra relativistic heavy-ion experiments.\\
\begin{table}
\begin{center}
\caption{Predictions of total mean multiplicity of charged hadrons produced in heavy-ion collision experiments.}
\resizebox{0.5\textwidth}{!}{
\begin{tabular}{cccc} \hline \hline 
{\bf Experiment}                                          &  {\bf LHC}     & {\bf CBM I}                           & {\bf CBM II}      \\ \hline \\
Colliding Nuclei                                    & $Pb-Pb$  & $Au-Au$                         & $Au-Au$      \\ \\
 $\sqrt{s_{NN}}$ (in GeV)                                & $2760$   & $3.97$ ($E_{lab}\approx 8 A$)  & 8.1  ($E_{lab}\approx 35 A$)\\ \\
$\langle n_{ch}\rangle^{min. bias}_{our ~model}$    & $8895$   & $328.75$                       & $670.8$ \\ \\
$\langle n_{ch}\rangle^{central}_{our ~model}$       & $14377$  & $380.8$                       & $837$ \\ \\
$\langle n_{ch}\rangle^{central}$~\cite{[armesto],[busza]} & $\approx 13124$ ($N_{part} =386$)              & --  & --\\ \\
 $\langle n_{\pi}\rangle^{central}_{HSD}$~\cite{[cbm],[cassing]} &--                            & $\approx 308$ & $\approx 650$\\ \hline
          
\end{tabular}
}
\end{center}
\end{table}
\noindent
{\bf Acknowledgments}\\

 PKS is grateful to the University Grants Commission (UGC), New Delhi for providing a research grant. BKS also sincerely acknowledges the Indian Space Research Organization (ISRO), India for providing the financial support.

%

\end{document}